\documentclass[aps,twocolumn,amsmath,amssymb,floatfix]{revtex4}
\usepackage{graphicx}
\usepackage{dcolumn}
\usepackage{bm}
\usepackage{amsfonts}
\usepackage{graphicx}
\usepackage{dcolumn}
\usepackage{bm}
\usepackage{amsmath}
\usepackage{amssymb}

\begin{document}
\title{Effects of disorder and internal dynamics on vortex wall propagation}
\author{Hongki Min$^{1,2}$}
\author{Robert D. McMichael$^{1}$}
\author{Michael J. Donahue$^{3}$}
\author{Jacques Miltat$^{1,2,4}$}
\author{M. D. Stiles$^{1}$}
\affiliation{
$^{1}$Center for Nanoscale Science and Technology, National Institute of Standards and Technology, Gaithersburg, Maryland 20899-6202, USA\\ 
$^{2}$Maryland NanoCenter, University of Maryland, College Park, Maryland 20742, USA\\ 
$^{3}$Mathematical and Computational Sciences Division, National Institute of Standards and Technology, Gaithersburg, Maryland 20899-8910, USA\\ 
$^{4}$Laboratoire de Physique des Solides, Universit\'{e} Paris Sud, CNRS, UMR 8502, 91405 Orsay, France
}
\date{\today}

\begin{abstract}
Experimental measurements of domain wall propagation are typically interpreted by comparison to reduced models that ignore both the effects of disorder and the internal dynamics of the domain wall structure. Using micromagnetic simulations, we study vortex wall propagation in magnetic nanowires induced by fields or currents in the presence of disorder. We show that the disorder leads to increases and decreases in the domain wall velocity depending on the conditions. These results can be understood in terms of an effective damping that increases as disorder increases. As a domain wall moves through disorder, internal degrees of freedom get excited, increasing the energy dissipation rate.
\end{abstract}

\maketitle
\normalsize

The dynamics of magnetic domain wall structures driven by fields or currents is a subject of practical importance related to possible schemes for nanoscale magnetic memory \cite{parkin_patent,parkin2008,hayashi2008} and logic \cite{allwood2002,allwood2005} devices. In these devices, information is encoded in the magnetic domains separated by domain walls and the stored information is manipulated by domain wall motion driven either by fields or currents. 

Experimentally, domain wall dynamics have been studied by the magneto-optical Kerr effect \cite{atkinson2003,vernier2004,beach2005,beach2006,yang2008}, resistance measurements using the giant magnetoresistance effect \cite{ono1999,grollier2003} or the anisotropic magnetoresistance effect \cite{tsoi2003,hayashi2006,hayashi2007}, and real-space magnetic imaging by magnetic force microscopy \cite{yamaguchi2004} or spin-polarized scanning electron microscopy \cite{klaui2005,uhlig2009}. Typical experiments measure a domain wall displacement and a time interval which are used to infer an average velocity. Interpretations of these results typically ignore the effects of disorder. Real samples, however, display thickness fluctuations and grain structure, and contain impurities and other defects. 

The consequences of disorder on domain wall motion have been studied theoretically in several limits. Micromagnetic simulations show that sample edge roughness can enhance domain wall propagation in a Ni$_{80}$Fe$_{20}$ wire \cite{nakatani2007,martinez2007}.  The dynamics of domain walls in the presence of a single pinning potential \cite{tatara2006} or array of pinning potentials \cite{ryu2009} show the existence of a threshold field or current to depin domain walls trapped by the pinning potentials. Moreover, domain wall creep \cite{metaxas2007} is common for distributed disorder at finite temperatures. 

In this Letter, we describe micromagnetic simulations of domain wall propagation induced by fields or currents in the presence of disorder throughout the film. Our results indicate that disorder, which exists inevitably in real experiments, affects domain wall dynamics in a way that can be interpreted as an enhancement of the effective damping.  This increase is significant enough that it should affect the interpretation of most domain wall experiments. Our work adds important considerations to the extraction from experiment of the intrinsic damping constant and the closely related nonadiabatic spin-transfer torque parameter.

\begin{figure}
\includegraphics[width=3.05in]{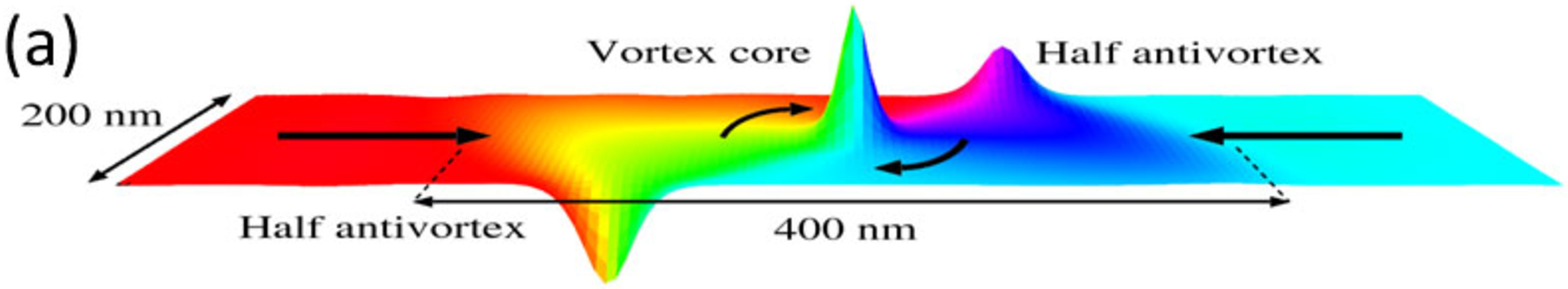}
\includegraphics[width=2.9in]{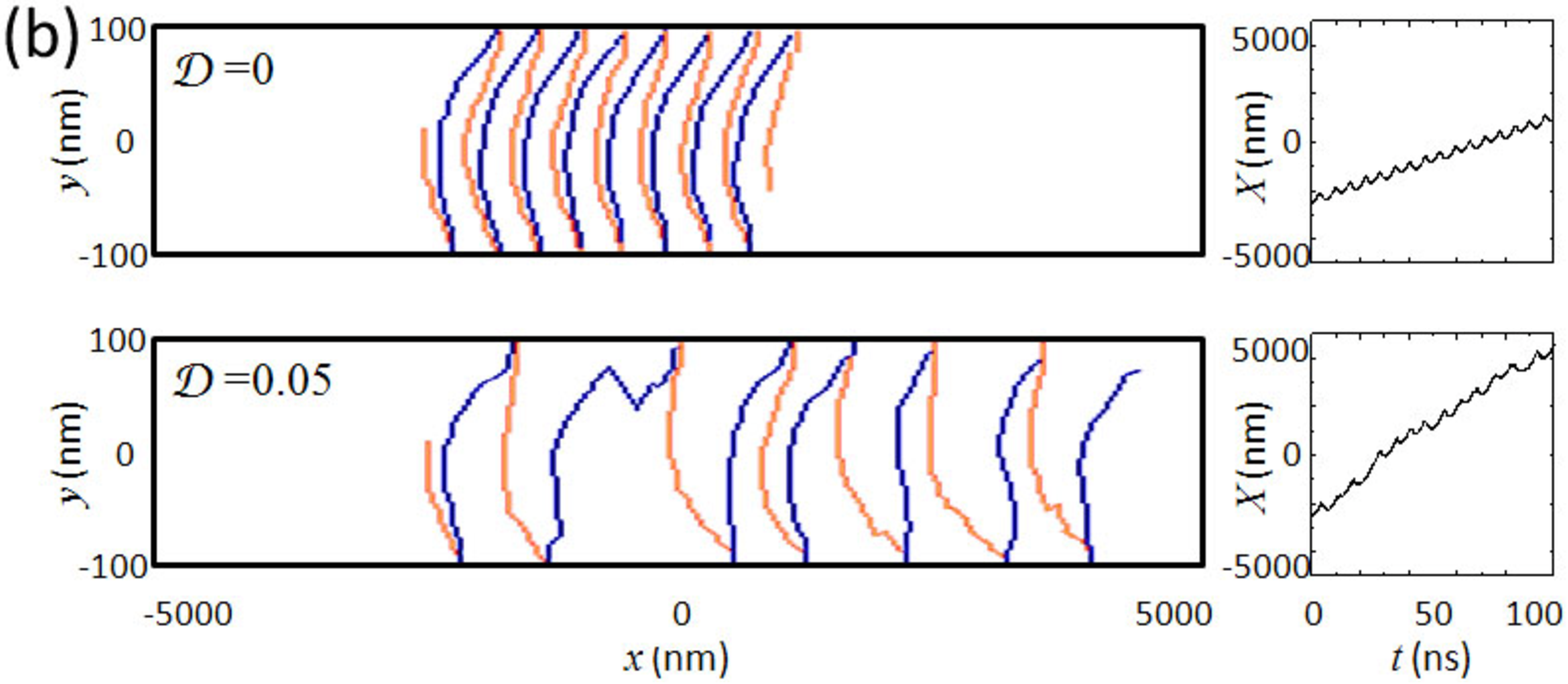}
\caption{(color online). (a) A typical vortex wall structure in a wire with 200 nm width and 20 nm thickness. The color indicates the in-plane angle of the magnetization, and the arrows indicate the approximate magnetization direction. (b) Schematic trajectories of field-induced vortex wall propagation in Ni$_{80}$Fe$_{20}$ film for $\mu_0 H$=3 mT above the critical field along the $x$ direction with disorder ${\cal D}$=0 and 0.05. Here the total simulation time is 100 ns. Points that the vortex core pass through are black (dark blue) or gray (orange) depending on whether the vortex core has its magnetization into or out of the plane. Insets show the domain wall displacement as a function of time.}
\label{fig:trajectory_H}
\end{figure}

Magnetization dynamics in the presence of a spin current can be described by an extended Landau-Lifshitz-Gilbert equation \cite{zhang2004,thiaville2005}
\begin{eqnarray}
\label{eq:LLG}
\dot{\bf M}&=&\gamma {\bf H}_{\rm eff}\times {\bf M} 
+\alpha\hat{\bf M}\times\dot{\bf M}
\nonumber\\
&&
-\left({\bf v}_{\rm s} \cdot \nabla\right){\bf M} 
+\beta\hat{\bf M}\times\left({\bf v}_{\rm s} \cdot \nabla\right){\bf M},
\end{eqnarray}
where ${\bf H}_{\rm eff}$ is the effective magnetic field including the external, exchange, demagnetization, and anisotropy fields, $\gamma$ is the gyromagnetic ratio, $M_{\rm s}$ is the saturation magnetization, $\hat{\bf M}$$=$${\bf M}/M_{\rm s}$, and $\alpha$ is the Gilbert damping constant.  The coupling between the current and the magnetization is characterized by two parameters.  The first is the velocity ${\bf v}_{\rm s}$$=$$P{\bf J}g\mu_{\rm B}/(2eM_{\rm s})$, where $P$ is the polarization of the current, ${\bf J}$ is the current density, $g$ is the Land\'{e} factor, $\mu_{\rm B}$ is the Bohr magneton, and $e$ is the (negative) charge of the electron.  The second parameter is the nonadiabatic spin-transfer torque parameter $\beta$.

In magnetic nanowires, the magnetization tends to point along the wires.  Domain walls form between domains of oppositely directed magnetization with demagnetization fields giving them complicated structures depending on the wire geometry \cite{mcmichael1997,nakatani2005}. The domain wall structure of interest here is a vortex wall, in which the magnetization in the wall rotates around a vortex core and points out of the plane of the wire at the core region.  This magnetization configuration is illustrated in Fig.~\ref{fig:trajectory_H}(a).  The configuration also contains two half antivortices on each of the edges of the wire.  

When a magnetic field is applied to a vortex wall, the vortex core displaces to the side of the wire.  If the field is below a value called the Walker breakdown field \cite{schryer1974}, the core then moves steadily along the wire.  If the field is above the breakdown field, the vortex core collides with the edge of the wire, reverses its magnetization, and moves to the other side.  The vortex core moves along the wire as it collides with both edges, as illustrated in the first panel of Fig.~\ref{fig:trajectory_H}(b).  Similar motion results when a current is applied to the wire.  

The motion of domain walls is frequently studied in models which adopt a reduced description of domain wall structures in terms of a limited number of collective coordinates \cite{schryer1974,thiaville2005,li2004,clarke2008,guslienko2002}. These models, however, ignore the additional degrees of freedom that may be excited during domain wall motion and further ignore the degree to which the excitation of these additional degrees of freedom change in the presence of disorder.  These effects are captured in micromagnetic simulations.

We compute domain wall motion through numerical solution of Eq.~(\ref{eq:LLG}) using the Object Oriented MicroMagnetic Framework (OOMMF) \cite{oommf}. We set up a Ni$_{80}$Fe$_{20}$ strip with 200 nm width, 20 nm thickness, and 5 nm cell size, and choose a long enough length to allow for subsequent domain wall propagation (typically from 10000 nm to 15000 nm).  In this geometry, vortex wall structures are formed as the ground state between head-to-head magnetic domains, as shown in Fig.~\ref{fig:trajectory_H}(a).  For material constants, we use the saturation magnetization $M_{\rm s}$=800 kA/m, exchange stiffness constant $A$=13 pJ/m, and damping constant $\alpha$=0.01.  In order to remove finite size effects, we add two features to the simulations. First, when we truncate the infinite wire we are modeling, there are unwanted fringing fields at the ends of the finite segment. We compensate these fields with static magnetic fields. Second, we include absorbing boundary conditions \cite{berkov2006} to remove spin waves reflected back to the computational region.

We model thickness fluctuations by varying the saturation magnetization $M_{\rm s}$ \cite{msfluct} but keeping the geometry uniform for simplicity.  We choose a spatial correlation length of 10 nm \cite{compton2006} and characterize the disorder as the ratio of fluctuation standard deviation to the saturation magnetization, ${\cal D}$=$\sqrt{\left<(M({\bf r})-M_{\rm s})^2\right>}/M_{\rm s}$.  We limit the size of the fluctuations to ensure that the magnetization stays positive.  To test this disorder model, we made comparisons (not shown) with recent gyration experiments \cite{compton2006}, which show a factor of 2$-$3 variation in resonance frequency as a vortex is scanned over a disk-shaped sample.  We find that a disorder value of 0.05 gives roughly the same variation in simulations with similar length scales. We tested different disorder models, such as random anisotropy directions with enhanced anisotropy constants, and found that our main results remain unaltered.

Figure \ref{fig:trajectory_H}(b) shows the effect of disorder on field-induced vortex wall propagation above $H_{\rm W}$. In the absence of disorder (upper panel), the vortex wall moves regularly from side to side of the wire switching the sign of core magnetization each time as it collides into the boundary. In the presence of disorder (lower panel), the vortex wall propagates both irregularly and faster.  The wall moves faster because disorder complicates the wall motion, increasing the fluctuations of the magnetization, and hence enhancing the total rate of energy dissipation into the lattice. 

\begin{figure}
\includegraphics[width=2.9in,height=1.43in]{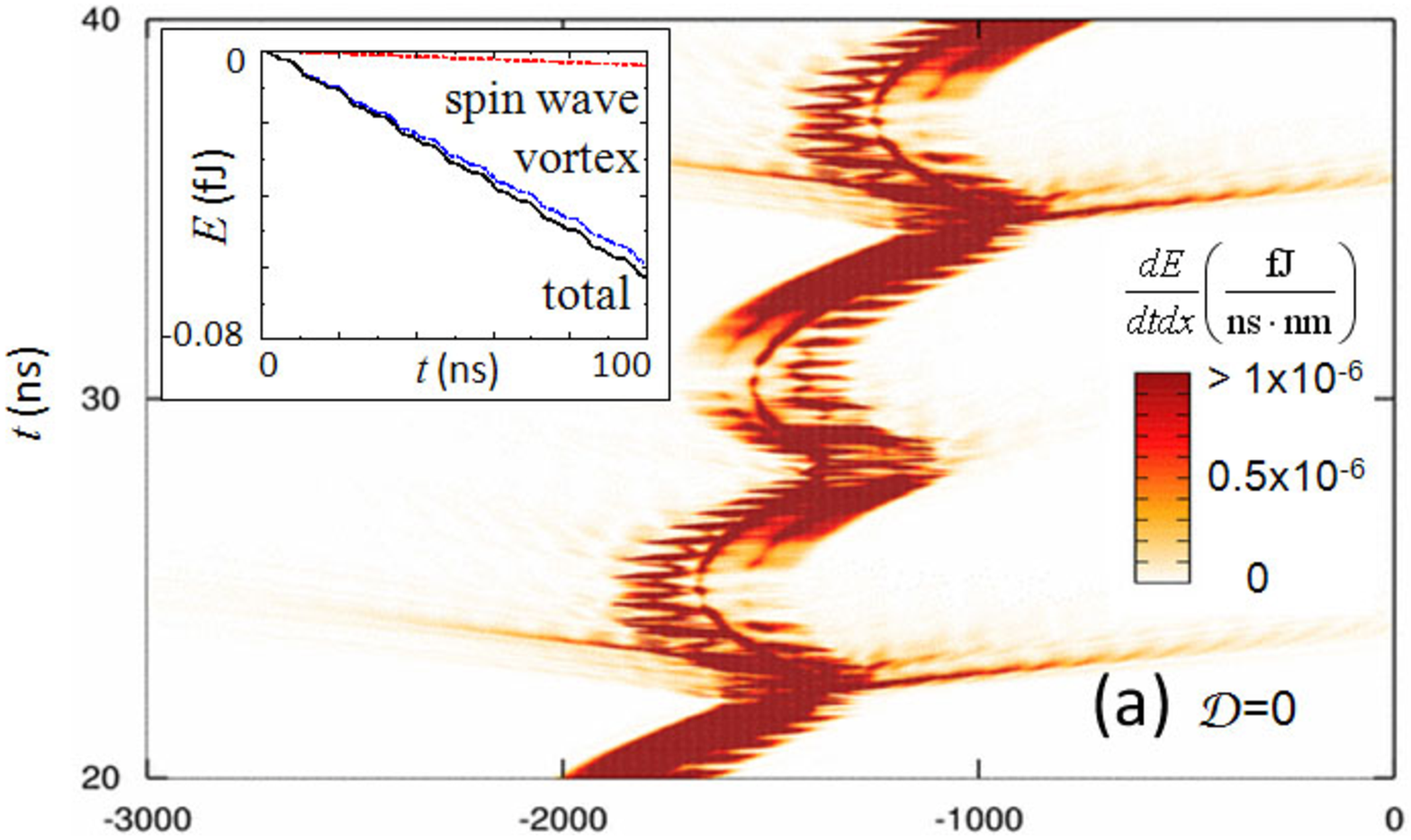}
\includegraphics[width=2.9in,height=1.57in]{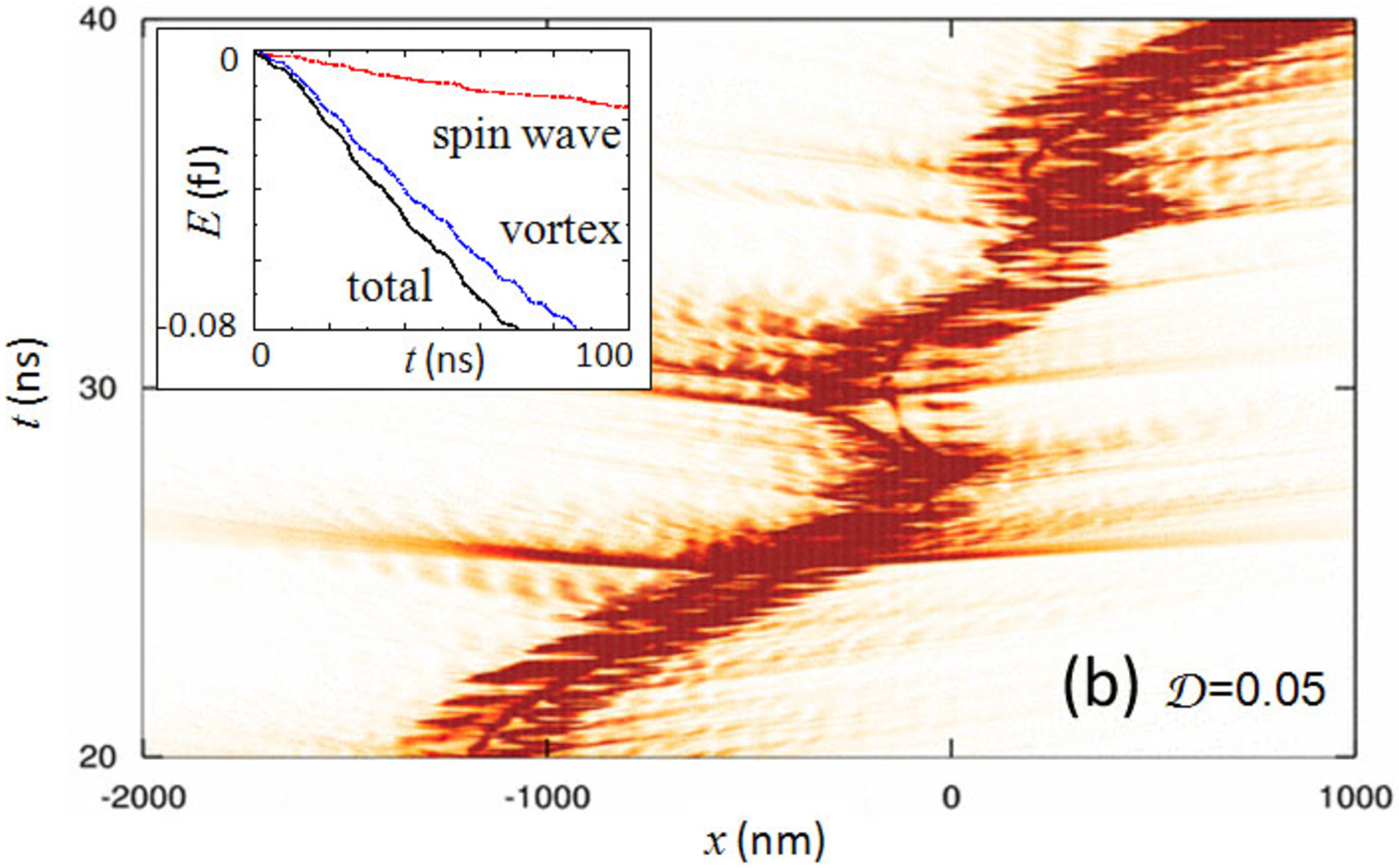}
\caption{(color online). Energy dissipation rate along $x$ for $\mu_0 H$=3 mT with disorder (a) ${\cal D}$=0 and (b) ${\cal D}$=0.05. Insets show contributions from spin wave and vortex to the energy dissipation rate, which were obtained by separating regions near a vortex core with a diameter of 400 nm, as shown in Fig.~\ref{fig:trajectory_H}(a).}
\label{fig:dissipation_H3}
\end{figure}

Figure \ref{fig:dissipation_H3} shows the local energy dissipation rate summed over the width and thickness of the wire as a function of the position along the wire ($x$ axis) and time ($y$ axis).  At each time, there are peaks in the dissipation rate where the magnetization changes rapidly in time as the domain wall moves, particularly around the vortex core and around each of the half antivortices. This motion is an example of an internal degree of freedom that is left out of a description of the domain wall in terms of collective coordinates.  The straight lines running left or right and slightly up indicate the emission of spin wave packets when the vortex core collides with the boundary. This emission is much stronger for collisions with one wire edge than the other because the collisions with the edges are not symmetric. For the field values considered, the magnetization in vortex walls rotates with a fixed handedness around the vortex core, which when combined with the applied field breaks the symmetry of the vortex relative to the two edges.  Because of this asymmetry, the core has a significantly higher velocity approaching one edge than it does approaching the other.

The inset of Fig.~\ref{fig:dissipation_H3}(a) shows that most of the energy dissipation occurs in a 400 nm wide region around the vortex core rather than through spin wave emission.  In the presence of disorder, both energy dissipation centered around the core and through spin wave emission increase, as shown in the inset of Fig.~\ref{fig:dissipation_H3}(b).  Spin wave emission is not just associated with collisions with the boundary, but apparently also with motion of the core through patches of strong disorder.  However, the dominant contribution to the increased rate of energy dissipation occurs in the localized region of the domain wall itself indicating the increase in the excitement of the internal degrees of freedom of the domain wall.  Note that the enhanced damping presented here is quite different than the two-magnon contribution to the linewidth as measured in ferromagnetic resonance.  We tested this by carrying out simulations of ferromagnetic resonance without vortex wall structures and found a much smaller enhancement of the effective damping.

\begin{figure}
\includegraphics[width=2.9in,height=1.7in]{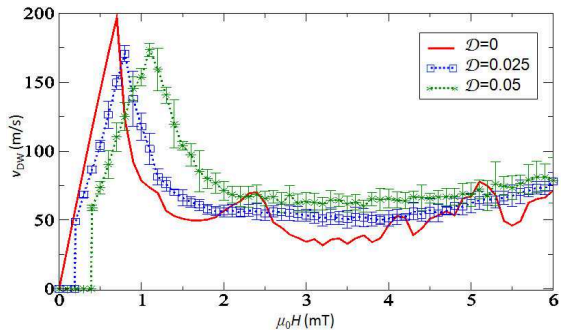}
\caption{(color online). Domain wall velocity as a function of applied field for disorder ${\cal D}$=0, 0.025, and 0.05.  Error bars indicate 1 standard deviation statistical uncertainty.}
\label{fig:dis_V_H}
\end{figure}

Figure \ref{fig:dis_V_H} shows the domain wall velocity as a function of applied field for disorder ${\cal D}$=0, 0.025, and 0.05. Here the domain wall velocity is estimated by ensemble averages of up to 40 samples with different realizations of the disorder. The disorder suppresses or enhances the domain wall velocity depending on the field range. At low enough fields, the domain walls are pinned in the presence of disorder. Note that the Walker breakdown field ($H_{\rm W}$$\approx$0.7 mT in the absence of disorder) itself is increased by the disorder.

The results in Fig.~\ref{fig:dis_V_H} show that even in the absence of disorder, vortex wall motion is complicated.  The curve for no disorder, ${\cal D}$=0, shows the expected linear rise as the field increases up to the breakdown field, then the subsequent decrease and increase as the field increases further.  However, the domain wall velocity as a function of field also shows additional peaks above the breakdown field, $H$$>$$H_{\rm W}$.  Increasing disorder suppresses  the peaks in the velocity curve, which may be the reason that they are not seen in experiments.  We observe that the spacing of peaks increases with increased intrinsic damping constant, and with the increased sample width. These results suggest that the origin of the peaks is a resonance between periodic collisions and the internal excitations of vortex wall structures.  It would be interesting if magnetic nanowires could be fabricated with sufficiently low disorder to observe such features.

The results in Fig.~\ref{fig:dis_V_H} can be understood in terms of an increase in effective damping parameter due to disorder. For field driven motion, the velocity depends strongly on the energy dissipation rate because a translation of the domain wall along the wire reduces the Zeeman energy.  If the internal energy of the wall is not changed, the wall can only move as this Zeeman energy is dissipated into the lattice.  In the reduced models mentioned above, we expect $v_{\rm DW}$$\sim$$\alpha$ for $H$$\gg$$H_{\rm W}$  because the energy dissipation rate and drift velocity of the wall increase  as $\alpha$ increases, while $v_{\rm DW}$$\sim$$1/\alpha$ for $H$$<$$H_{\rm W}$ because, as $\alpha$ increases, the displacement of the core toward the sample edge decreases, and the drift velocity and the energy dissipation rate decrease \cite{thiaville2005,li2004}. We also note that $H_{\rm W}$$\sim$$\alpha$. In the presence of disorder, the domain wall velocity increases as disorder increases for $H$$>$$H_{\rm W}$, while for $H$$<$$H_{\rm W}$, the domain wall velocity decreases, exactly as would be expected for an increase in the effective damping parameter. 

\begin{figure}
\includegraphics[width=2.95in,height=1.7in]{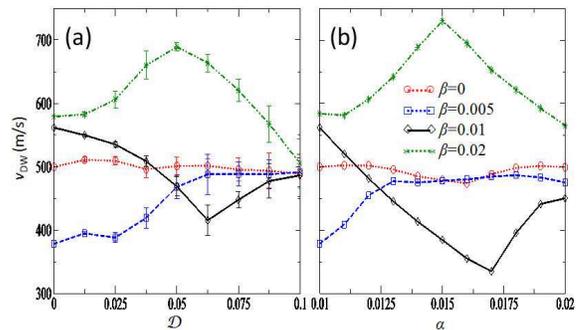}
\caption{(color online). Domain wall velocity (a) as a function of disorder and (b) as a function of the damping constant for $J$=$2$$\times$$10^{13}$ A/m$^2$ with $\beta$ as a parameter.}
\label{fig:dis_V_J}
\end{figure}

In the case of current-induced domain wall propagation, the results can also be interpreted by an enhanced effective damping. While we expected that the effective value of $\beta$ would increase as well, we find that changing $\alpha$ alone provides the best explanation of the results. To see this behavior, we compare calculations of the domain wall velocity as a function of disorder with calculations without disorder but increasing damping constant, both with fixed $\beta$.  Figure \ref{fig:dis_V_J}(a) shows the domain wall velocity as a function of disorder ${\cal D}$ for $J$=$2$$\times$$10^{13}$ A/m$^2$, which is above the critical current $J_{\rm c}^{\beta=0}$$\approx$$0.8$$\times$$10^{13}$ A/m$^2$.  Figure \ref{fig:dis_V_J}(b) shows the domain wall velocity as a function of the damping constant $\alpha$ with the same applied current density. As the disorder increases, the variation of the domain wall velocity increases or decreases depending on $\beta$ showing a clear resemblance to the results with the enhanced damping constant in the absence of disorder.

\begin{figure}
\includegraphics[width=3in,height=1.7in]{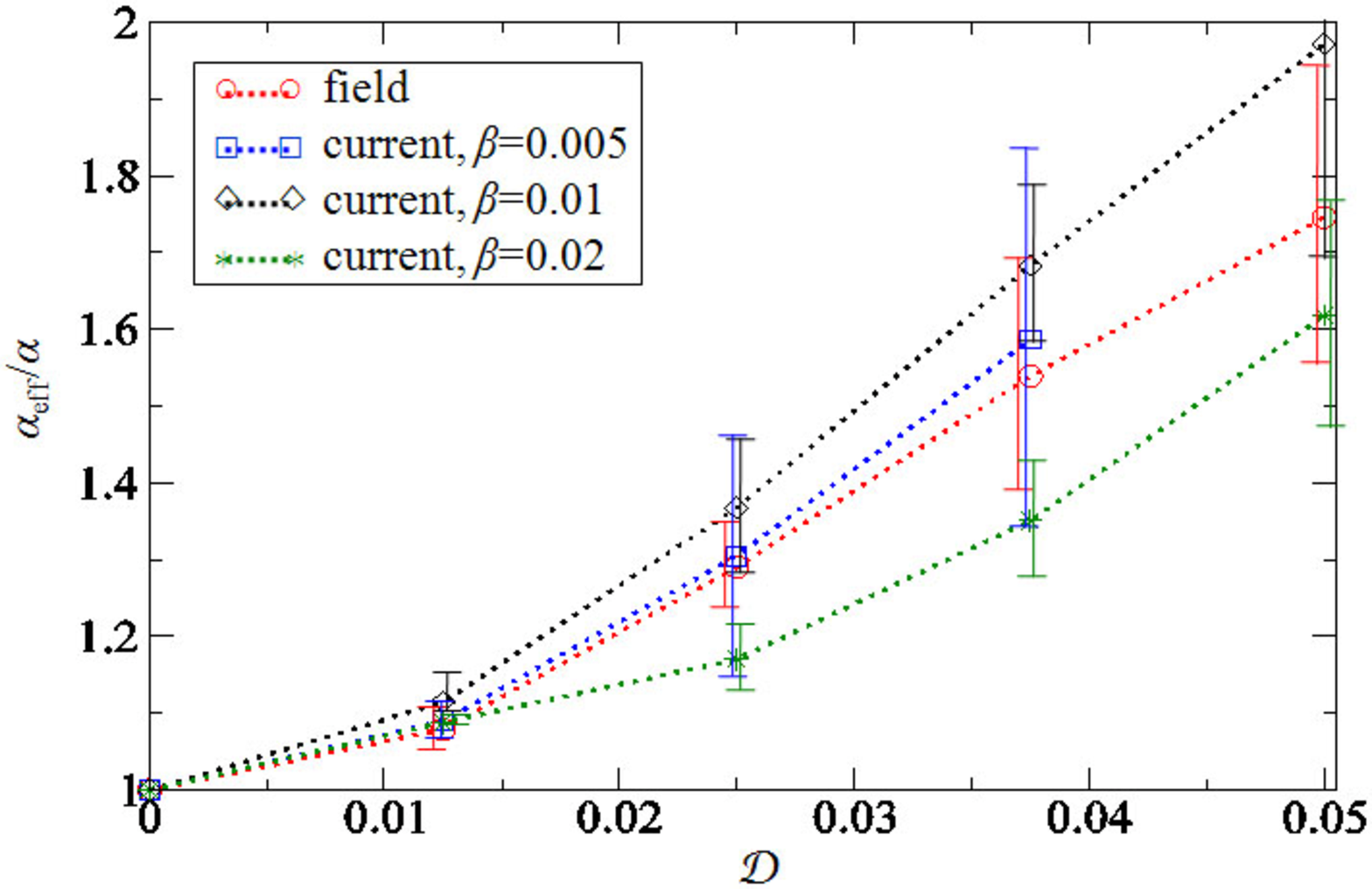}
\caption{(color online). Effective damping $\alpha_{\rm eff}$ as a function of disorder ${\cal D}$ for field-induced and current-induced vortex wall propagation.
}
\label{fig:dis_a}
\end{figure}

We compute the disorder dependence of effective damping by fitting the domain wall velocity in the linear low field and low current regime, as shown in Fig.~\ref{fig:dis_a}. We point out that the actual values of the disorder-enhanced damping rate depend on various factors such as the type of the domain wall structures, the type of disorder, geometry of samples, and material properties. Several experiments would test the results of our calculations. One possible experiment is to measure the domain wall velocity with a disorder introduced in a controlled manner.  Another possible experiment would be the vortex gyration in a single pinning potential in which the enhanced damping could be measured by comparing the spectrum between free and trapped regimes of vortex gyration.

In summary, we have demonstrated that disorder affects domain wall dynamics significantly and that the effective damping is increased by disorder and internal excitations of the domain wall structure.  From this work, we conclude that damping constants inferred from domain wall motion measurements are effective rather than intrinsic values, which are enhanced by the disorder in a sample.  These results suggest that caution is necessary in extracting fundamental parameters from  domain wall motion measurements.

The work has been supported in part by the NIST-CNST/UMD-NanoCenter Cooperative Agreement. The authors thank J. J. McClelland, K. Gilmore, and June W. Lau for their valuable comments.



\begin{thebibliography}{999}
\bibitem{msfluct}
Locally varying $M_{\rm s}$ gives the lowest order variation in both the magnetostatic interaction and $v_{\rm s}$ consistent with thickness fluctuations.  We ignore changes in the exchange coupling, which tests indicate does not strongly affect our results.
\end{thebibliography}
\end{document}